\documentclass[natbib]{nature_fig}
\usepackage{amssymb}
\usepackage[utf8]{inputenc}
\usepackage{textgreek}
\usepackage{amsmath}
\usepackage{txfonts}
\usepackage{threeparttable}

\usepackage{hyperref}
\hypersetup{
colorlinks=true,
linkcolor=blue,
citecolor=blue,
urlcolor=blue
}

\usepackage{graphicx}
\usepackage{caption}
\usepackage{url}
\usepackage{lineno}
\usepackage{subcaption}

\usepackage[numbers,square]{natbib}
\usepackage{tikz}
\usetikzlibrary{shapes.geometric, arrows}

\usepackage{lscape}
\usepackage{longtable}
\usepackage{multirow} 
\usepackage{threeparttable}
\usepackage{threeparttablex}
\usepackage{pdflscape}

\newcommand{\apjl}{Astrophys. J. Lett.}   
\newcommand{\apjs}{Astrophys. J. Suppl. Ser.}   
\newcommand{\ao}{Appl. Opt.}   
\newcommand{\aap}{Astron. Astrophys.}   

\newcommand{\gray}{$\gamma$-ray~}
\newcommand{\grays}{$\gamma$-rays~}

\usepackage[utf8]{inputenc}

\title{LHAASO Detection of Ultra-High-Energy Gamma-Ray Emission toward the Giant Molecular Clouds}

\author{
Zhen Cao$^{1,2,3}$,
F. Aharonian$^{3,4,5,6}$,
Y.X. Bai$^{1,3}$,
Y.W. Bao$^{7}$,
D. Bastieri$^{8}$,
X.J. Bi$^{1,2,3}$,
Y.J. Bi$^{1,3}$,
W. Bian$^{7}$,
A.V. Bukevich$^{9}$,
C.M. Cai$^{10}$
% ...
(The LHAASO Collaboration)
Y.H. Yu$^{37}$

$^{1}$ Key Laboratory of Particle Astrophysics \& Experimental Physics Division \& Computing Center, Institute of High Energy Physics, Chinese Academy of Sciences, 100049 Beijing, China\\
$^{2}$ University of Chinese Academy of Sciences, 100049 Beijing, China\\
$^{3}$ TIANFU Cosmic Ray Research Center, Chengdu, Sichuan,  China\\
$^{4}$ University of Science and Technology of China, 230026 Hefei, Anhui, China\\
$^{5}$ Yerevan State University, 1 Alek Manukyan Street, Yerevan 0025, Armeni a\\
$^{6}$ Max-Planck-Institut for Nuclear Physics, P.O. Box 103980, 69029  Heidelberg, Germany\\
$^{7}$ Tsung-Dao Lee Institute \& School of Physics and Astronomy, Shanghai Jiao Tong University, 200240 Shanghai, China\\
$^{8}$ Center for Astrophysics, Guangzhou University, 510006 Guangzhou, Guangdong, China\\
$^{9}$ Institute for Nuclear Research of Russian Academy of Sciences, 117312 Moscow, Russia\\
$^{10}$ School of Physical Science and Technology \&  School of Information Science and Technology, Southwest Jiaotong University, 610031 Chengdu, Sichuan, China\\
% ...
$^{37}$ School of Physics, Henan Normal University, Xinxiang 453007, Henan, China\\ 
}

\date{ 2025}
\begin{document}

\maketitle

\clearpage
\linenumbers 
\begin{abstract}
The $\gamma$-ray from Giant molecular clouds (GMCs) is regarded as the most ideal tool to perform in-situ measurement of cosmic ray (CR) density and spectra in our Galaxy. We report the first detection of $\gamma$-ray emissions in the very-high-energy (VHE) domain from the five nearby GMCs with a stacking analysis based on a 4.5-year $\gamma$-ray observation with the Large High Altitude Air Shower Observatory (LHAASO) experiment. The spectral energy distributions derived from the GMCs are consistent with the expected $\gamma$-ray flux produced via CR interacting with the ISM in the energy interval 1 - 100 $~\rm$ TeV. In addition, we investigate the presence of the CR spectral `knee' by introducing a spectral break in the $\gamma$-ray data. While no significant evidence for the CR knee is found, the current KM2A measurements from GMCs strongly favor a proton CR knee located above 0.9$~\rm$ PeV, which is consistent with the latest measurement of the CR spectrum by ground-based experiments. 
\end{abstract}

The current paradigm of cosmic rays (CRs) suggests that the majority of CR flux, up to the so-called knee around $10^{15}$~eV, originates from galactic sources, most likely supernova remnants (e.g., \cite{Drury2012}). It is generally assumed that CRs, during their propagation through the interstellar magnetic fields, become well mixed, resulting in a relatively uniform CR density across the Galactic disk. The CR density measured locally near Earth is often considered representative of the average CR density across the galaxy, referred to as the "sea" of galactic CRs. However, while large-scale (kpc) variations in CR density are minimal, significantly small-scale variations, especially near young CR accelerators, cannot be excluded.

It is thus not guaranteed that the locally measured CR component represents the entire galactic population of relativistic particles. In particular, the local CR flux might be dominated by a few nearby sources. In this regard, gamma-ray observations can provide unique information on the spatial distributions of CRs in our Galaxy. 

Giant Molecular Clouds (GMCs) are massive reservoirs of gas and dust, with masses typically around $10^5$ M$_\odot$ and sizes ranging from tens to hundreds of parsecs. These clouds are primarily composed of molecular hydrogen and are some of the densest structures in the ISM, with densities several orders of magnitude higher than the average in the solar vicinity. Due to their substantial mass and density, GMCs are critical regions for studying CR interactions and their effects on the ISM. They serve as natural laboratories to observe the impact of CRs on the surrounding matter and to probe variations in CR density throughout the galaxy.

Thus, the measurement of gamma-ray emissions from GMCs is a promising way to probe the "sea" of CRs and their distribution throughout the galaxy. In this regard, GMCs also serve as ideal targets for studying CR propagation. Studies such as those using Fermi-LAT data have employed this method to investigate cosmic-ray propagation, notably in regions like GMCs in the Gould Belt (\cite{Ackermann2012ApJb, Ackermann2012ApJa, Nerinov12, Yang13}). These works calculated gamma-ray emissivities and molecular mass conversion factors to gain insights into the CR spectrum across different regions of the galaxy. These analyses have similarly focused on high-latitude clouds, where the gamma-ray spectra derived from massive clouds align with local CR measurements, confirming that CRs in the energy range of 10-100 GeV are consistent with direct measurements from experiments such as PAMELA \cite{Adiani2011PhRvL} and AMS \cite{Aguilar2015PhRvL}. However, some recent studies, such as Ref.\cite{Baghmanyan2020ApJ}, have reported discrepancies, finding higher-than-expected gamma-ray emission in certain clouds, suggesting possible variations in the CR `sea'.  In higher energy, HAWC has recently reported the observations towards several nearby GMCs, the derived upper limit is consistent with the predicted gamma-ray emissivities assuming the same CR density as local measurements \cite{hawc_gmc2021ApJ}. These findings underscore the importance of continued study of gamma-ray emission to explore variations in the CR sea and the influence of local sources.  With these observations, it becomes possible to probe CR density variations at GeV and potentially TeV energies, using gamma rays as tracers of CR interactions in GMCs. Close-by GMCs, particularly those associated with the star formation complexes of the Gould Belt, present valuable targets for gamma-ray studies, allowing researchers to further refine our understanding of CR propagation throughout the galaxy. 

\section{The $\gamma$-ray flux of the giant molecular clouds} 
\begin{figure*}
    \centering
    \includegraphics[width=0.39\linewidth]{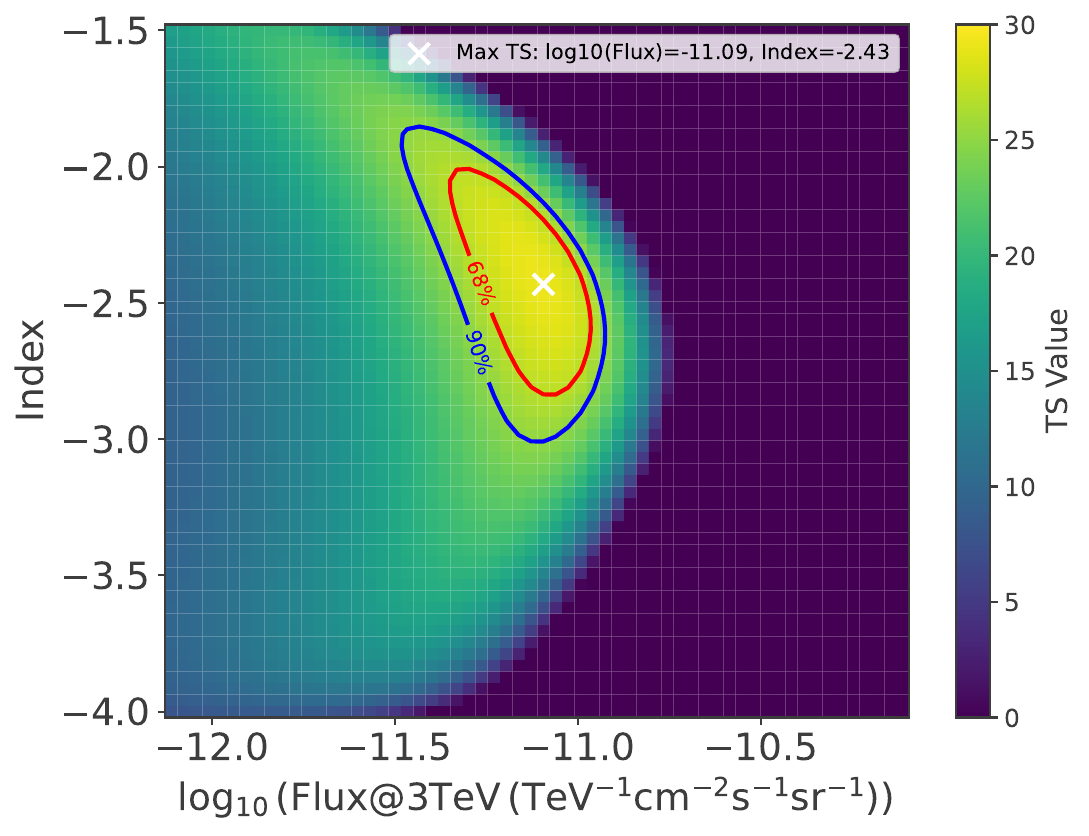}
    \includegraphics[width=0.41\linewidth]{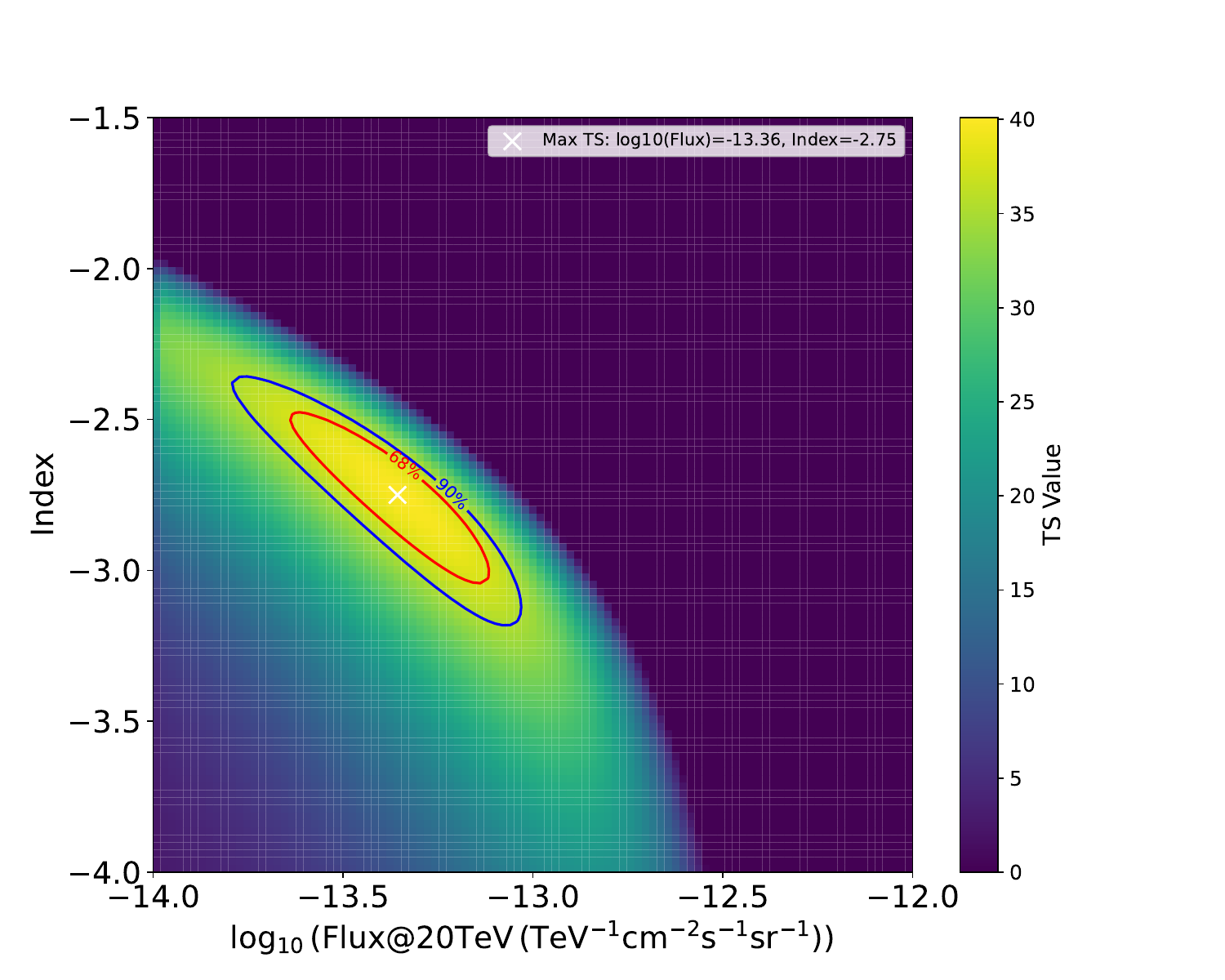}
    \caption{Stacked TS profile for GMC sample containing 5 sources. The spectrum of GMCs is assumed to follow a power-law distribution. The TS value is color-coded for each flux and index combination. The `x' sign shows the maximum value. The two solid contours represent 68$\%$ and 90$\%$ confidence level.}
    \label{fig:km2a-pl-TSprofile}
\end{figure*}

We selected five massive clouds - Hercules, Monoceros, Orion, Perseus, and Taurus, which are in the field of view of LHAASO with a high galactic latitude ($\lvert b \rvert >$ 5$^\circ$) (see Method \ref{sect:gmc}). We analyze the WCDA data with a total live time of 979 days and the KM2A data with a total live time of 1733 days to search for the $\gamma$-ray emission from those five GMCs (see Method \ref{data_analysis}). As shown in the Supplementary Figure. \ref{fig:km2a-skymap}, we produced significant sky maps of those five GMC regions listed in Supplementary Table \ref{tab:GMC_info},  and found no significant excess at each cloud. The TS value for each GMC  produced by 3D Likelihood analysis is too low ($<$ 25) to give a significant detection. A stacking technique was adopted for searching for the weak $\gamma$-ray emission from the GMCs. We stacked their individual TS profiles in the  1~\rm TeV-30~\rm TeV energy range with WCDA data and 25~\rm TeV-1~\rm PeV energy range with KM2A data to search for $\gamma$-ray emission and explore average properties. Note that, in the stacking analysis, the contribution from each cloud is weighted according to its average gas column density, normalized to a reference value of $1\times10^{22}$ cm$^{-2}$. This weighting reflects the physical expectation that the $\gamma$-ray flux is approximately proportional to the gas column density in hadronic interaction scenarios. As shown in the  Figure. \ref{fig:km2a-pl-TSprofile}, significant gamma-ray emission is detected from the five clouds with a maximum TS value of 30.0 for WCDA data and 40.1 for KM2A data, corresponding to the best-fit result of flux@ 3 ~\rm TeV = $(8.04^{+1.56}_{-2.44})\times10^{-12}$ ph TeV$^-1 $s$^{-1}$ cm$^{-2}$ sr$^{-1}$ with an index of -2.43$^{+0.28}_{-0.26}$ for WCDA,  and flux@20~\rm TeV = $(4.36^{+2.09}_{-1.35})\times10^{-14}$ ph TeV$^-1$s$^{-1}$ cm$^{-2}$ sr$^{-1}$ with an index = -2.75$^{+0.20}_{-0.13}$ for KM2A. The differential flux and the best-fitted spectral index derived from the stacking analysis are shown in Figure. \ref{fig:km2a-pl-expected}.   Meanwhile, we also analyzed the combined KM2A and WCDA datasets and found no significant reduction in the resulting TS value compared to those obtained individually from the KM2A and WCDA data (with a difference of 3). The spectral index obtained from the joint analysis is -2.83$^{+0.08}_{-0.08}$. 

Since we combined the \gray undetected objects, a genuine question arises about the possibility of stacking the false \gray emission rather than the radiation originating from the actual signal produced via interactions between cosmic rays and the ISM. To evaluate the contribution from random background fluctuations, we selected several random positions with low gas column density (see Method \ref{data_analysis}). As shown in Supplementary Figure.\ref{fig:disk_template}, the stacking result yields a maximum TS value below 2, confirming that the contribution from random background fluctuations to the stacked $\gamma$-ray emission from the clouds is negligible. To further verify whether the observed excess is indeed associated with the molecular cloud, we performed a stacking analysis using a disk template in the same regions. The test shows no significant excess, supporting the interpretation that the observed signal is correlated with the molecular cloud.

\section{Comparison with the predicted gamma-ray flux and constraining to the CR spectrum}  
\begin{figure*}
    \centering
    \includegraphics[width=0.85\linewidth]{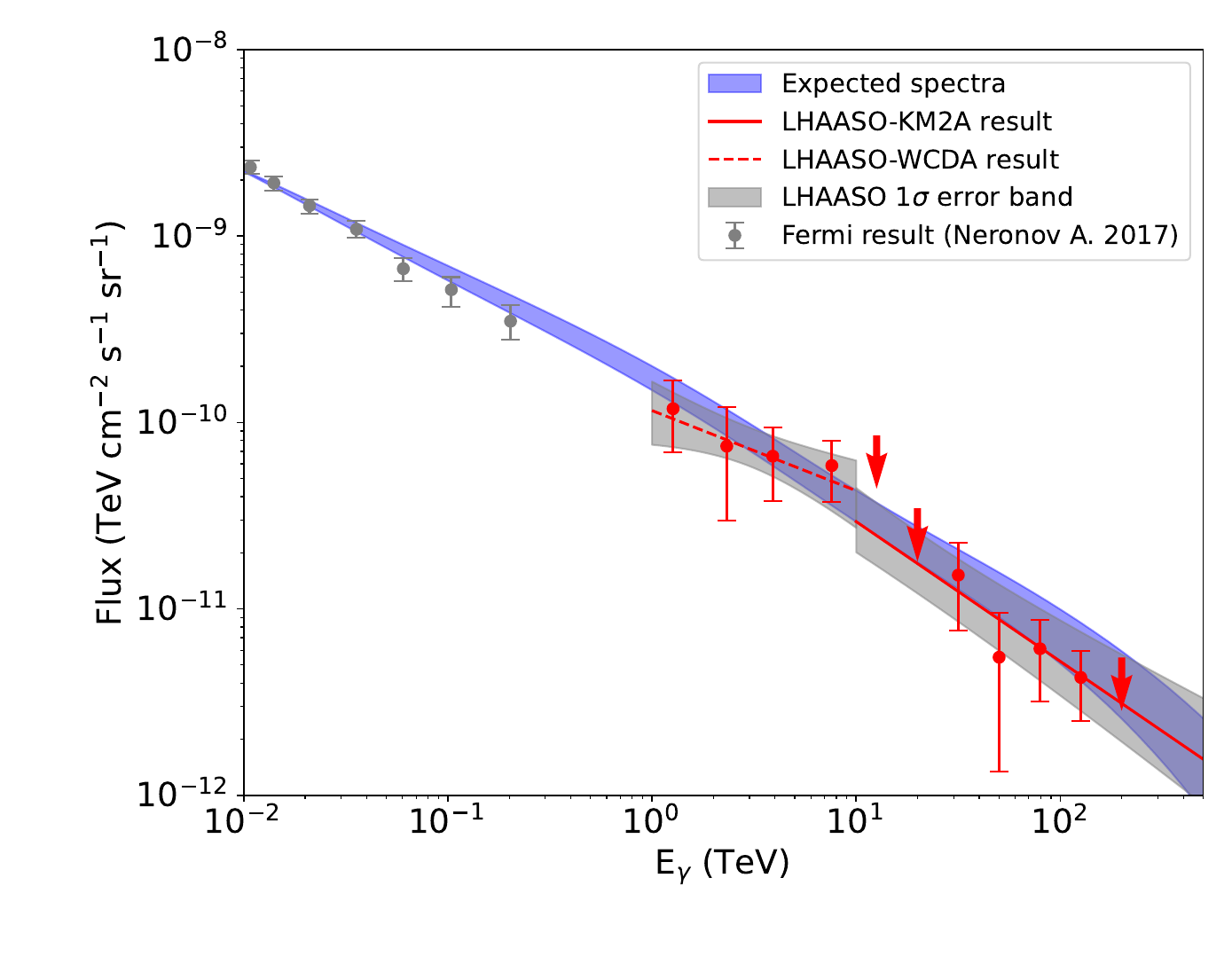}
    \caption{The SED of the GMC as measured by LHAASO-KM2A. The differential flux has been normalized to a column density of $1\times10^{22}$ cm $^{-2}$ according to the column density of each GMC. The red solid (dashed ) line shows the best-fit power-law function of the KM2A (WCDA) data, and the gray shaded band is the $\pm1\sigma$ statistical uncertainty. The measurement by Fermi  (gray points) is shown for comparison, which has been scaled based on the column density and distance of the cloud \cite{2017A&A...606A..22N}.  The blue shaded band represents the expected \gray flux produced via CR interacting with ISM, assuming a spectrum of CR which coincides with that used in the \cite{2023KM2A_diffuse}. }
    \label{fig:km2a-pl-expected}
\end{figure*}

Due to the proximity of the GMCs in this study, we assume that the CR spectra in the GMCs are identical to the local measurement. We compare the LHAASO measurements of the GMCs emission with the predictions of hadronic interactions between CRs and the ISM.  We calculated the \gray emissivities by using the {\it AAFRAG} package \citep{Kachelries2019CoPhC} and the gamma-ray production cross section parametrized in Ref.\cite{Kafexhiu2014PhRvD}, and get the expected gamma-ray flux. The CR proton and helium spectra are summarized in Ref.\cite{2023KM2A_diffuse}. The `high' and `low' \gray flux are from the `high' and `low' representative of CR spectra in Ref.\cite{2023KM2A_diffuse}, respectively. As shown in the Figure. \ref{fig:km2a-pl-expected}, we found that the measured \grays flux assuming a single power law is in good agreement with the predicted flux.  

We note that the CR spectra have several fine structures from about $100~\rm GeV$ to $100~\rm TeV$ before the `knee', but such slight changes in the CR spectra only have minor effects on the secondary \gray spectra and can hardly be traced by current \gray measurements. However, all current direct measurements have revealed a significant change in the power law index in the knee, and the position of the `knee' would have a significant impact on the secondary \gray spectra. On the other hand, the \gray production is dominated by the CR protons, since the nuclear interactions which produce secondary \grays scale with kinetic energy per nucleon.  In the current understanding of the acceleration mechanism the acceleration is rigidity related, if we assume that the `knee' mainly reflects the maximum acceleration limit of the CR accelerators in our Galaxy, near the `knee' the production of  \grays should be dominated by CR protons due to the larger maximum kinetic energy per nucleon assuming the same maximum rigidity for all nuclei species.  Thus, the \gray measurement can provide unique information on the position of the proton `knee'.   While the direct measurement of CR by the ground-based instruments can only reconstruct the total kinetic energy.  In this work, we neglect the fine structures in the CR spectra before the `knee' and use a broken power law spectrum to model the secondary \gray spectrum below and above the `knee', the position of the break in \gray spectrum $E_{break}$ is determined directly by the position of the proton `knee'. 

\begin{figure*}
    \centering
    \includegraphics[width=0.85\linewidth]{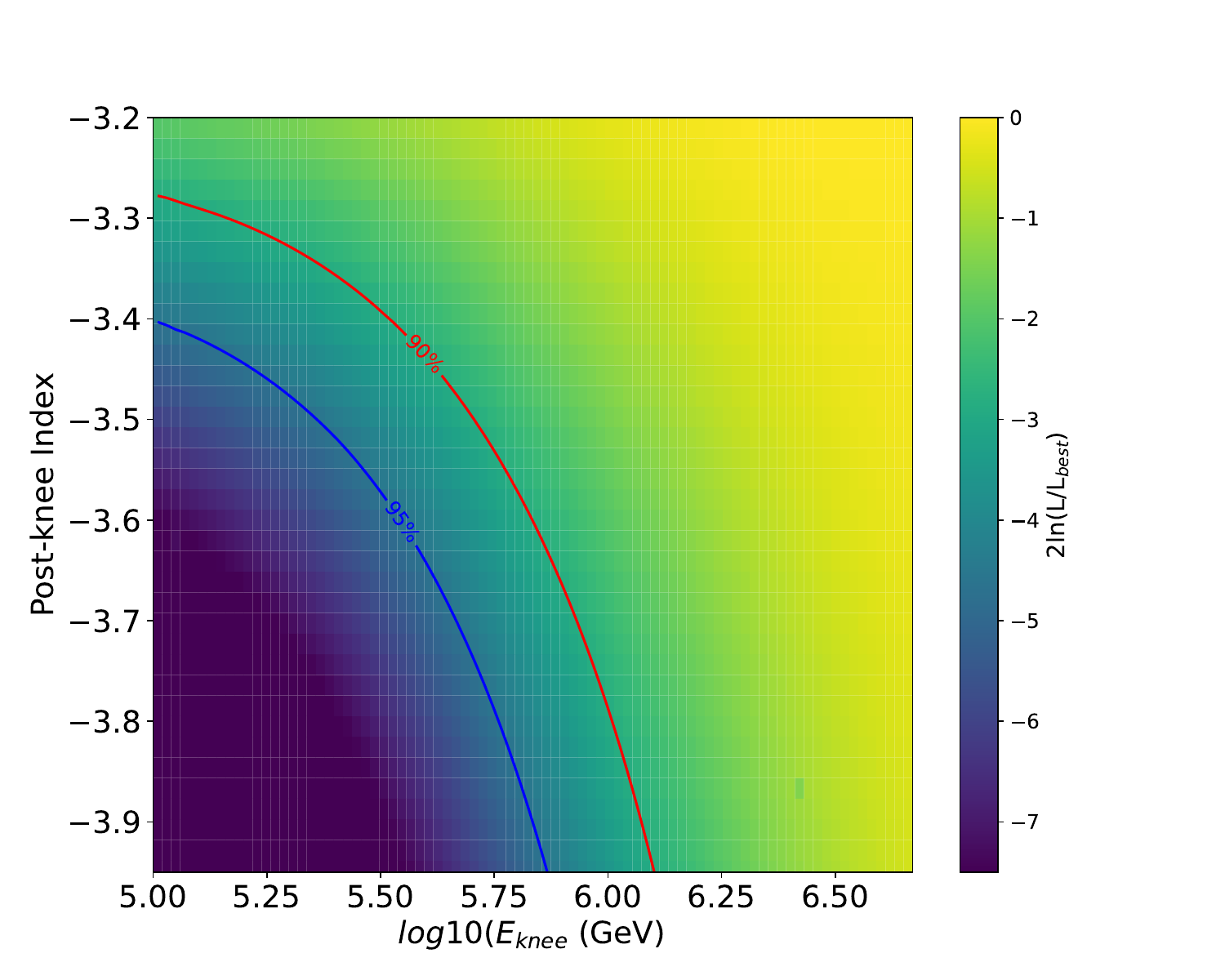}

    \caption{
    Stacked TS profile for GMCs obtained by fitting KM2A data with the predicted gamma-ray spectrum using the  $AAFRAG$ package, assuming various values of $E_{knee}$ of cosmic rays and the post-knee index. The spectral shape was fixed during the fit, with the flux normalization left free. Significance contours are overlaid on the plot showing the 90$\%$ and 95$\%$ confidence levels. 
    }
    \label{fig:km2a-sbpl-TSprofile}
\end{figure*}

We used a broken power law spectrum as shown in Eq. \ref{equ2} of Method \ref{data_analysis} to represent the \gray spectrum.  The inclusion of such an energy break cannot improve the likelihood fit, which implies that the current data cannot actually detect the `knee'. To see the constraints on $E_{knee}$ from $E_{break}$, we calculated the expected gamma-ray flux assuming different $E_{knee}$. We used the CR proton and helium spectra measured by DAMPE \citep{dampe_p, dampe_he} up to $100~\rm TeV$, then extrapolated to higher energy and added a spectral break by hand to represent the position of the `knee'. The smoothness of the spectral break in CR spectra is fixed to be $s=2$, but we varied the post-knee index from $3.2-4.0$. The position of the Helium `knee' is set to be twice that of the `knee' of protons, which is expected if we assume the `knee' reflects the acceleration limit and the acceleration is rigidity dependent.  We analyzed the KM2A data using the predicted gamma-ray spectra calculated with the {\it AAFRAG} package and obtained the TS stacked profile for the GMCs. The spectral shape was fixed during the fit, with the flux normalization left free.  As shown in the Figure. \ref{fig:km2a-sbpl-TSprofile}, it is evident that the `knee' below $0.9~\rm PeV$ produces the spectral break in gamma-rays at a significantly lower value than our  90\%  lower limit, with an assumed post-knee index of -3.7.  The current KM2A measurement on GMCs strongly favors a CR proton `knee' higher than $0.9~\rm PeV$, which is consistent with the latest LHAASO measurement of the CR proton spectrum \cite{LHAASO_proton}.   

\section{Conclusion}
In summary, we performed a stacking analysis on LHAASO data on the nearby GMCs and detected the gamma-ray emissions in the VHE-UHE domain with a significance of more than $5~\rm sigma$. The observed spectra in these regions are well described by a power-law model. Due to the lack of other potential gamma-ray emitters in these regions, we argue that the gamma-rays we detect are from the interaction of CRs with the ambient gases. Furthermore, the gamma-ray flux measured by LHAASO is consistent with predictions based on CR interactions with the ISM. 

The presence of the `knee' in the CR spectrum also implies a spectral break in the gamma-ray flux of these GMCs, which lies in the LHAASO KM2A energy range. We also performed a search for such spectral breaks in this study. The likelihood fitting was not improved by adding a spectral break in the gamma-ray spectrum, which indicated that we cannot detect the CR `knee' directly from the current LHAASO KM2A data. But in the context of likelihood fitting, we can derive a lower limit of the CR `knee' with the gamma-ray data. We found that the current KM2A observations favor a CR proton `knee' higher than $0.9~\rm PeV$ with an assumed post-knee index of -3.7. 

With the accumulation of LHAASO data, as well as the sky coverage of other nearby GMCs by the proposed SWGO \citep{2019swgo} experiments in the future, it is possible to detect the position of the CR `proton ' knee directly from the gamma-ray observations on these objects. In addition to the nearby GMCs, the remote molecular clouds can also be used to trace the CR spatial distributions in our Galaxy \citep{aharonian19, peron22}. This is also an important task in gamma-ray astronomy, since the CR density can vary significantly in different parts of our Galaxy\citep{strong98}, especially in the VHE/UHE domain\citep{semikoz23}. The recent observations on diffuse gamma-ray emissions in the Galactic plane also reveal a significant excess compared with the `standard' picture of CR propagation in our Galaxy \citep{2023KM2A_diffuse, zhang23}. The diffuse emission in the Galactic plane can only measure the average CR density in a specific line of sight.  In this regard, the detection of gamma-ray flux from dense molecular clouds in different parts of our Galaxy serves as the only way to measure the CR spatial distribution all over the Milky Way and is crucial in understanding the propagation and origin of the Galactic CRs.  

\clearpage
\clearpage

\clearpage 

\clearpage 
\clearpage

\part*{Methods}
\setcounter{section}{0} 
\section{The Giant Molecular Clouds}
\label{sect:gmc}
Using a similar method to that described by Ref.\cite{yang2014}, we selected five massive clouds identified in the CO galactic survey conducted by Ref.\cite{Dame2001}  with the CfA 1.2 m millimeter-wave Telescope. These five massive clouds, Hercules, Monoceros, Orion, Perseus, and Taurus, are in the field of view of LHAASO with a high galactic latitude ($\lvert b \rvert >$ 5$^\circ$). The observations with the Planck satellite provide opacity maps of these clouds, which can be used to derive precise information about the column density and masses of these GMCs. This information is shown in Table \ref{tab:GMC_info}. 
We used the Planck dust-opacity maps, which are sensitive to both molecular and atomic gas (including the "dark gas"), to trace the gas distributions and derive gas column density. The total gas column density can be calculated by relating it to the dust opacity, $\tau _{M} (\lambda)$, where $\lambda$ is the wavelength. According to equation (4) of Ref.\cite{planck2011}, the equation used to derive the gas column density is: 
\begin{equation}
    N_{H} = \tau_{M}(\lambda)[(\frac{\tau_{D}(\lambda)}{N_{H}})^{ref}]^{-1}
\end{equation}

The reference dust emissivity measured in low-$N_H$ regions at 353 GHz, $(\tau_D/N_H)^{ref}$, can be found from table 3 of Ref.\cite{planck2014}. The derived gas column density maps of these five GMCs are shown in Figure. \ref{fig:dustmap}. In constructing the templates of the GMCs based on the gas information, we account for the dependence of the detector’s angular resolution on both energy and zenith angle.

\section{LHAASO data analysis} 
\label{data_analysis}
\subsection{Likelihood analysis}
LHAASO, located at Haizi Mountain in China at an altitude of 4410~\rm m,  is dedicated to the $\gamma$-ray astronomy in the energy range from hundreds of GeV to 1~\rm PeV, as well as cosmic-ray physics.  It is composed of three sub-arrays: a 7800~\rm m$^2$ water Cherenkov detector array (WCDA) for TeV $\gamma$-ray astronomy, a 1.3~\rm km$^{2}$ array (KM2A) for $\gamma$-ray astronomy above 10 TeV, and 18 wide-field-of-view air Cherenkov/fluorescence telescopes (WFCTA) for cosmic ray physics in the energy range of 10~\rm TeV to 1~\rm EeV.  WCDA and KM2A have a wide field-of-view (FOV) of $\sim$2~\rm sr, which is essential for studying extended gamma-ray sources such as GMCs.   

The results presented here for WCDA were obtained using the full array configuration from March 5, 2021, to July 31, 2024, with a total live time of 979 days. The certain data selection and gamma-ray/background discrimination cut methods can be found in Ref.\cite{wcdacrab}. For the WCDA data, events are divided into five groups according to the number of triggered PMT units, referred to as $N_{hits}$, i.e., 100-200, 200-300, 300-500, 500-800, $>$ 800.   The KM2A data were collected by the half-array of KM2A from December 2019 to December 2020, three-quarters of KM2A from December 2020 to July 2021, and the whole KM2A array from July 2021 to December 2024. The total live time is 1733 days. The events of the KM2A  are divided into five groups per decade with a bin width of $\Delta log_{10} E$ =0.2 according to reconstructed energy \cite{KM2Acrab}.  The sky map in the celestial coordinate system (R.A. and Dec) is divided into a grid of 0.1$^\circ \times$0.1$^\circ$ and each cell is filled with the events according to their reconstructed arrival directions. The "direct integration method" \cite{background2004} is adopted to estimate the number of background events in each pixel to obtain the excess gamma-ray-induced showers.  

The analysis is performed using a 3-dimensional (3D) likelihood algorithm. The test statistic (TS) is used to compare the goodness of each hypothesis. It is defined as:
\begin{equation}
\mathrm{TS} = \frac{\mathcal{L}_{s+b}}{\mathcal{L}_{b}} = 
\frac{
\max \prod_i \mathrm{Poisson}\left(N_i^{\mathrm{obs}}, N_i^{\mathrm{exp1}}\right)
}{
\max \prod_i \mathrm{Poisson}\left(N_i^{\mathrm{obs}}, N_i^{\mathrm{exp0}}\right)
},
\end{equation}
where $\mathcal{L}_{s+b}$ is the maximum likelihood value for the hypothesis that includes both the sources and the background, while $\mathcal{L}_{b}$ is for the background-only hypothesis. `i' is the index of each bin with specific energy and position. $N^{\exp1}_{i}$ and $N^{\exp1}_{i}$ are the expected number of events in each bin of the alternative hypothesis and the null hypothesis, taking into account the detector response, respectively.

We first assume \gray flux follows a sample power-law shape for each cloud as $F = A\times(E/E_{0})^{-\alpha}$, where A is the differential flux at $E_{0}$, $E_{0}$ is the reference energy which is set to 3~\rm TeV for WCDA data and to 20 TeV for KM2A data, and $\alpha$ is the spectral index. Source templates are constructed by selecting high-density regions within each cloud, defined as those with column densities exceeding $5\times10^{21}$ cm$^{-2}$. For the Hercules cloud, a lower threshold of $2.5\times10^{21}$ cm$^{-2}$ is adopted, reflecting its overall lower column density. This corresponds to regions where the opacity is primarily dominated by the CO contribution, as discussed in Sec. 4 of Ref.\cite{planck2011}. To check for the presence of sources, we generated the TS maps. We assume a point source with a power-law spectrum with an index of 2.7 at each pixel, and use the maximum likelihood method to calculate the significance of each pixel of the sky map. According to the Wilks' Theorem \cite{Wilks:1938dza}, the TS value follows a $\chi^{2}$ distribution with one degree of freedom (d.o.f), and we take $\pm\sqrt{TS}$ as the significance.  

\subsection{Testing Spectral Break Predictions at the CR 'Knee' with KM2A Data}
The CR `knee' implies that the corresponding gamma-ray spectrum should also have a spectral break at about $100~\rm TeV$, which lies in the energy range of KM2A. Thus,  we performed the likelihood analysis by adding a spectral break to test such predictions with KM2A data. The spectra model is assumed to follow: 
\begin{equation}
    J(E) = \Phi_0 \times (\frac{E}{E_{0}})^{\gamma_{1}} (1+ (\frac{E}{E_{break}})^{s})^{(\gamma_2-\gamma_1)/s}
    \label{equ2}
\end{equation}
where $E_{break}$ corresponds to the break position, $\gamma_1$ and $\gamma_2$ are spectral asymptotic slopes before and after the break, $s$ is the sharpness parameter of the break.  Here, we fix the $s$ as 2 and $E_{0}$ as 20~\rm TeV. 
\subsection{Stacking technique} 
The stacking technique is often used to explore \gray average properties for astrophysical populations\cite{Paliya_2019,Ajello_2020,2023MNRAS.524.5854S}. We first generate a set of TS profiles for each cloud using a grid of photon flux values at 20~\rm TeV for KM2A data (ranging from 10$^{-15}$ to 10$^{-12}$ ph cm$^{-2}$ s$^{-1}$ sr$^{-1}$, divided into 300 logarithmic steps), or 3 ~\rm TeV for WCDA data (ranging from 10$^{-13}$ to 10$^{-10}$ ph cm$^{-2}$ s$^{-1}$ sr$^{-1}$, divided into 100 logarithmic steps ) and photon indices (ranging from -5 to -1 with a step size of 0.02). Note that, in the stacking analysis, the contribution from each cloud is weighted according to its average gas column density, normalized to a reference value of $1\times10^{22}$ cm$^{-2}$. This weighting reflects the physical expectation that the $\gamma$-ray flux is approximately proportional to the gas column density in hadronic interaction scenarios. Since the log-likelihood is additive in nature, we add the TS profiles of each cloud to create a combined profile, which represents the average spectral properties of all the GMCs. 

\subsection{Systematic uncertainties}
The `direct integration method' for background estimation may introduce large-scale structures due to the long integration time (10-hour window). Such structures may result from spurious anisotropy caused by the variations in the spatial distribution of the efficiency over the time window, as well as from the true cosmic-ray anisotropies. It is therefore necessary to evaluate the impact of these large-scale structures, introduced by the background estimation method, on the measurements. Using the method described in Ref.\citep{2023KM2A_diffuse}, we applied smoothing to the background sky map and incorporated measurements of large-scale cosmic-ray anisotropy obtained from actual observations \cite{Gao_anisotropy} to correct the background map. As shown in the Supplementary Figure. \ref{fig:Differentdata}, we found that this effect results in about 20$\%$ systematic uncertainty for the flux and 0.07 for the spectral index.    

\clearpage
\begin{addendum}

\item We would like to thank all staff members who work at the LHAASO site above 4400 meter above the sea level year round to maintain the detector and keep the water recycling system, electricity power supply and other components of the experiment operating smoothly. We are grateful to Chengdu Management Committee of Tianfu New Area for the constant financial support for research with LHAASO data. We appreciate the computing and data service support provided by the National High Energy Physics Data Center for the data analysis in this paper. This research work is supported by the following grants: The National Natural Science Foundation of China: No.12393854,  No. 12220101003, No.12175121, No.12275280, No.12393851, No.12393852, No.12393853,  No.12205314, No.12105301, No.12305120, No.12261160362, No.12105294, No.U1931201, No.12375107, and the CAS Project for Young Scientists in Basic Research (No. YSBR-061) and in Thailand by the National Science and Technology Development Agency (NSTDA) and the National Research Council of Thailand (NRCT) under the High-Potential Research Team Grant Program (N42A650868). 

\item[Author Contributions]
R.Z. Yang and Y.H. Yu initiated the project and led the data analysis and theoretical interpretation. Y.H. Yu analyzed the KM2A data, and W.Y. Cao analyzed the WCDA data. R. Zhang cross-checked the result. R.Z. Yang and J. Li performed the phenomenological interpretation. The whole LHAASO collaboration contributed to the publication, with involvement at various stages ranging from the design, construction and operation of the instrument, to the development and maintenance of all software for data calibration, data reconstruction and data analysis. All authors reviewed, discussed and commented on the present results and on the manuscript. 
\item[Corresponding author emails]

yangrz@ustc.edu.cn(R.Z. Yang);yuyanhong@htu.edu.cn (Y.H. Yu) ; caowy@ustc.edu.cn (W.Y. Cao); jianli@ustc.edu.cn (Jian Li); zhangrui@pmo.ac.cn (R. Zhang); 

\item[Competing interests]
The authors declare no competing interests.

\end{addendum}

\clearpage

\clearpage
\noindent {\bf References}
\bibliographystyle{naturemag}
\bibliography{refs}

\clearpage



\renewcommand{\figurename}{Supplementary Figure} 
\renewcommand{\tablename}{Supplementary Table} 
\setcounter{figure}{0}
\setcounter{table}{0} 

\begin{table*}[htbp!]
\centering
\begin{threeparttable}
\footnotesize 
\caption{\bf Properties of the GMCs}
\begin{tabular}{ccccc}
    \hline
    \hline
       GMC   &  Distance (pc) & R.A. (deg) & Decl. (deg) \\
    \hline
       Hercules  & 200$\pm$30&280.6&15.7\\
       Monoceros &830$\pm$83 &92.3&-6.8 \\
       Orion & 490$\pm$50 &87.3&-3.6\\
       Perseus &315$\pm$32 & 52.9&30.9 \\
       Taurus &140$\pm$30 &66.5&26.2\\
     \hline  
    \end{tabular}
\label{tab:GMC_info}
\end{threeparttable}
\end{table*}
\clearpage 
\begin{figure*}
    \centering

    \begin{subfigure}{0.35\linewidth}
        \includegraphics[width=\linewidth]{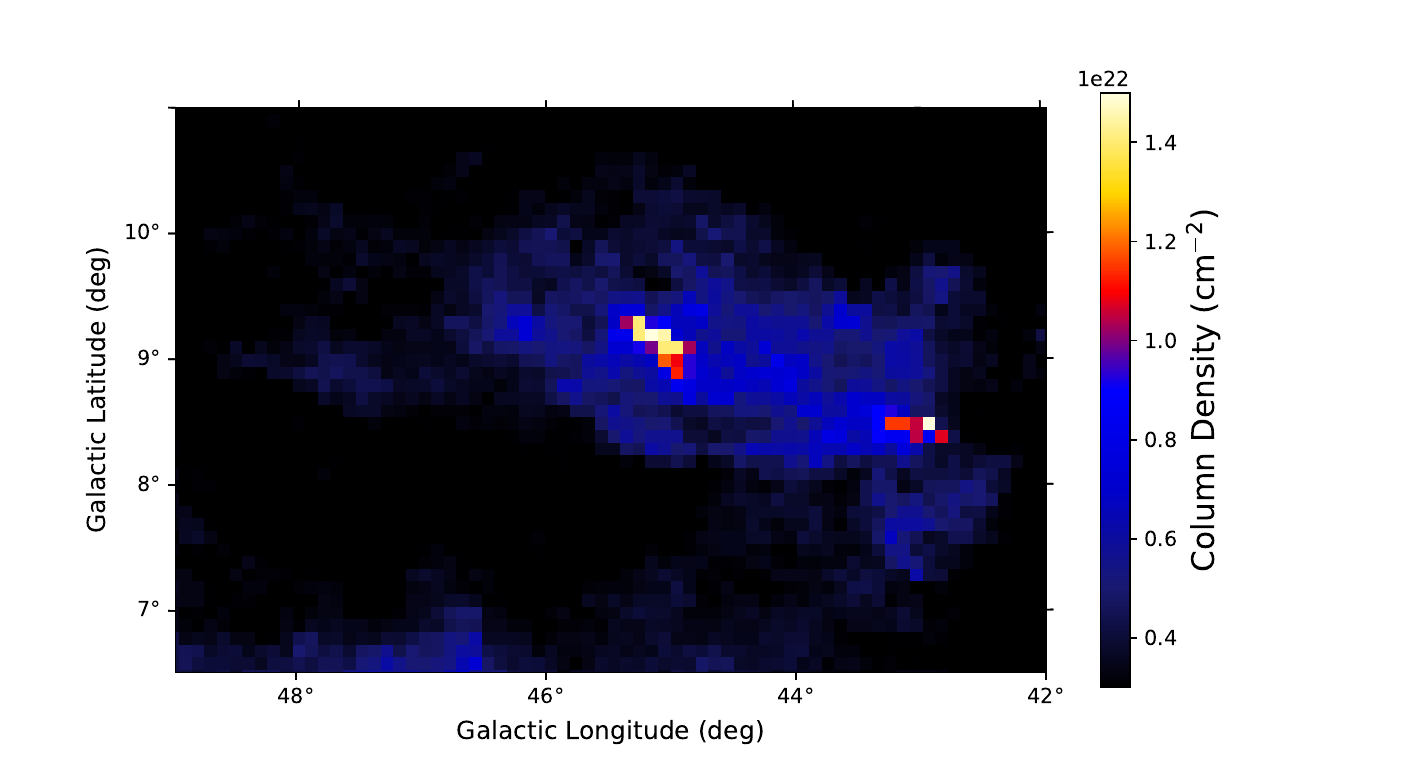}
        \caption{Hercules}
    \end{subfigure}
    \begin{subfigure}{0.35\linewidth}
        \includegraphics[width=\linewidth]{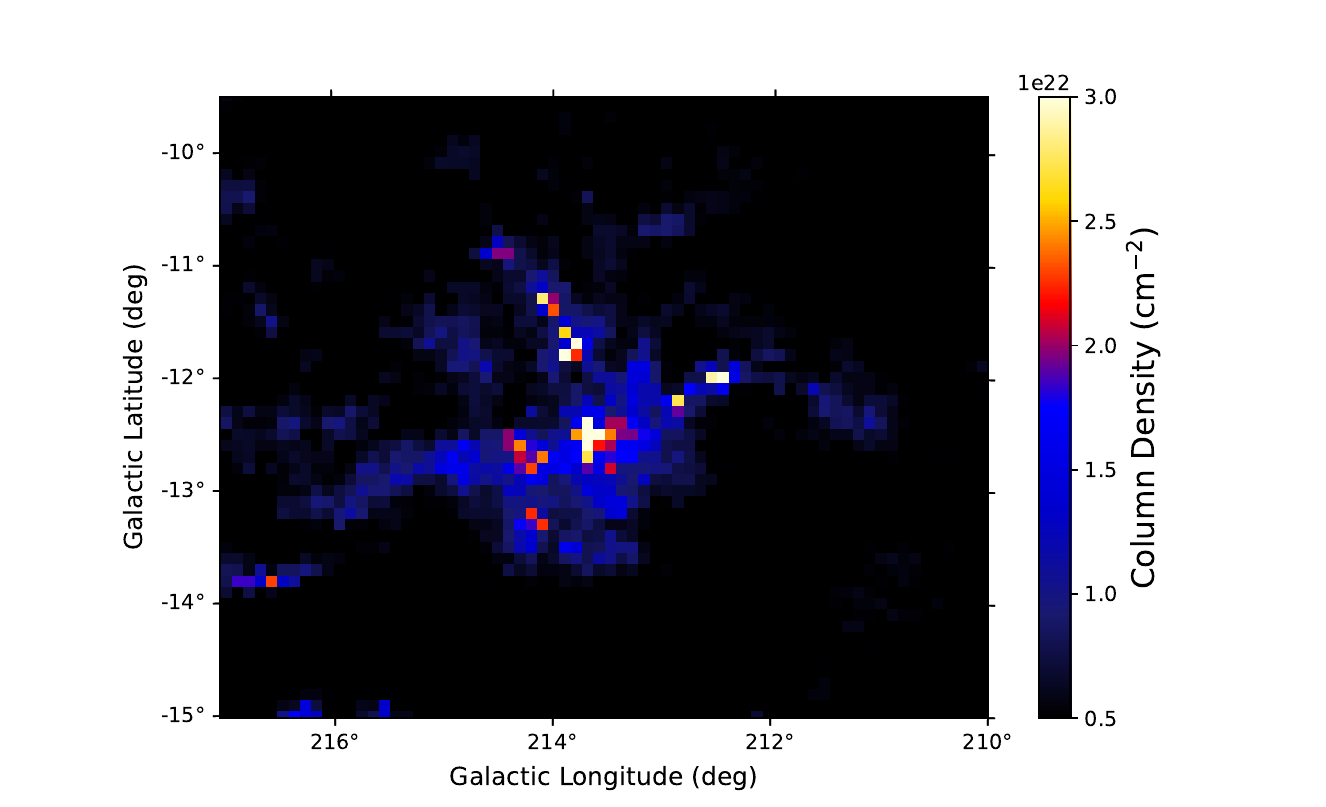}
        \caption{Monoceros}
    \end{subfigure}
    
    \begin{subfigure}{0.35\linewidth}
        \includegraphics[width=\linewidth]{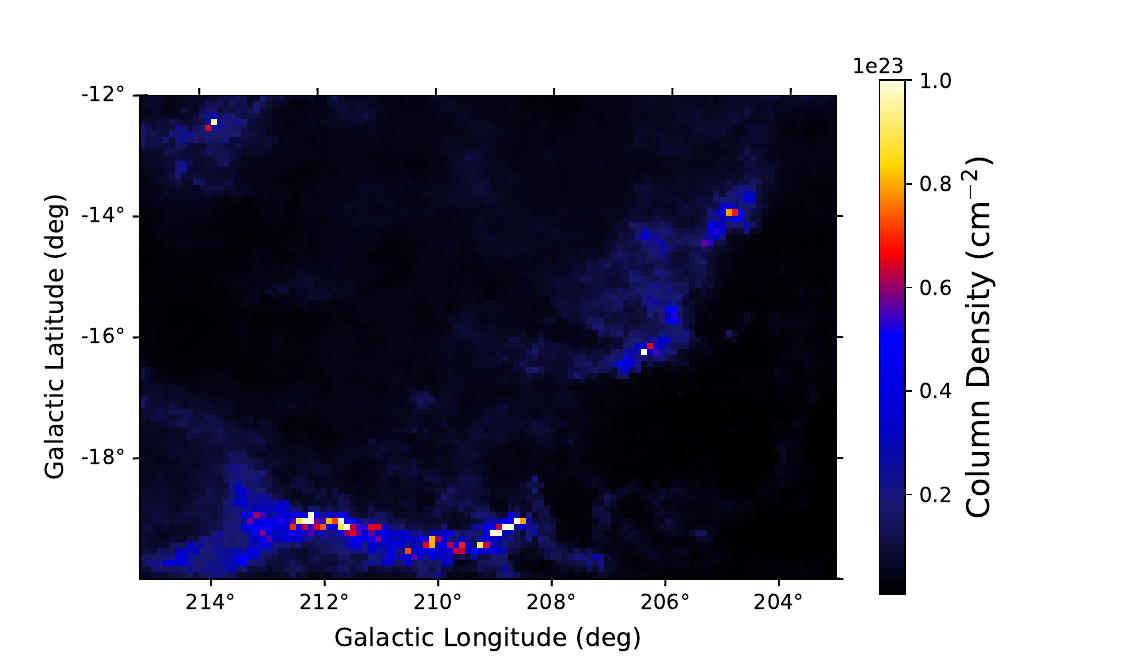}
        \caption{Orion}
    \end{subfigure}
    \begin{subfigure}{0.35\linewidth}
        \includegraphics[width=\linewidth]{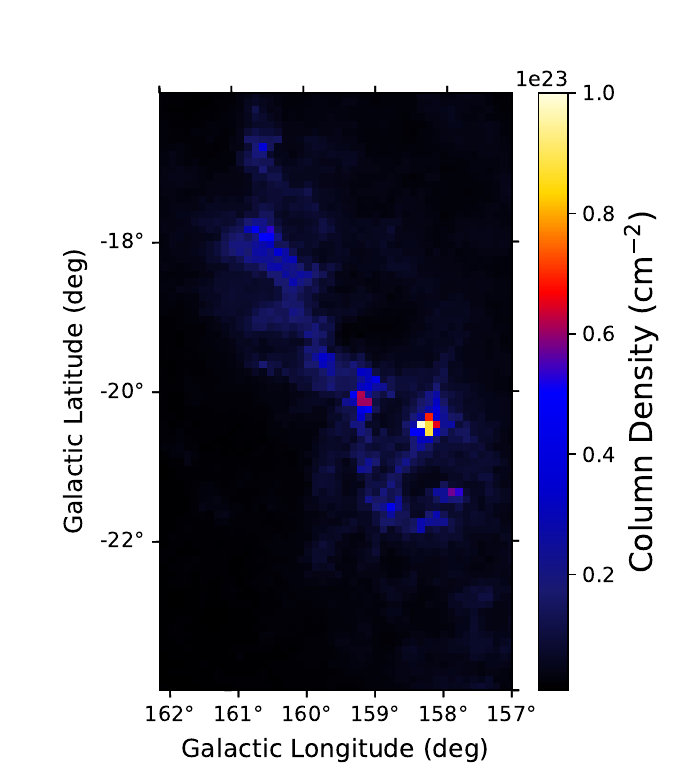}
        \caption{Perseus}
    \end{subfigure}
    
    \begin{subfigure}{0.35\linewidth}
        \includegraphics[width=\linewidth]{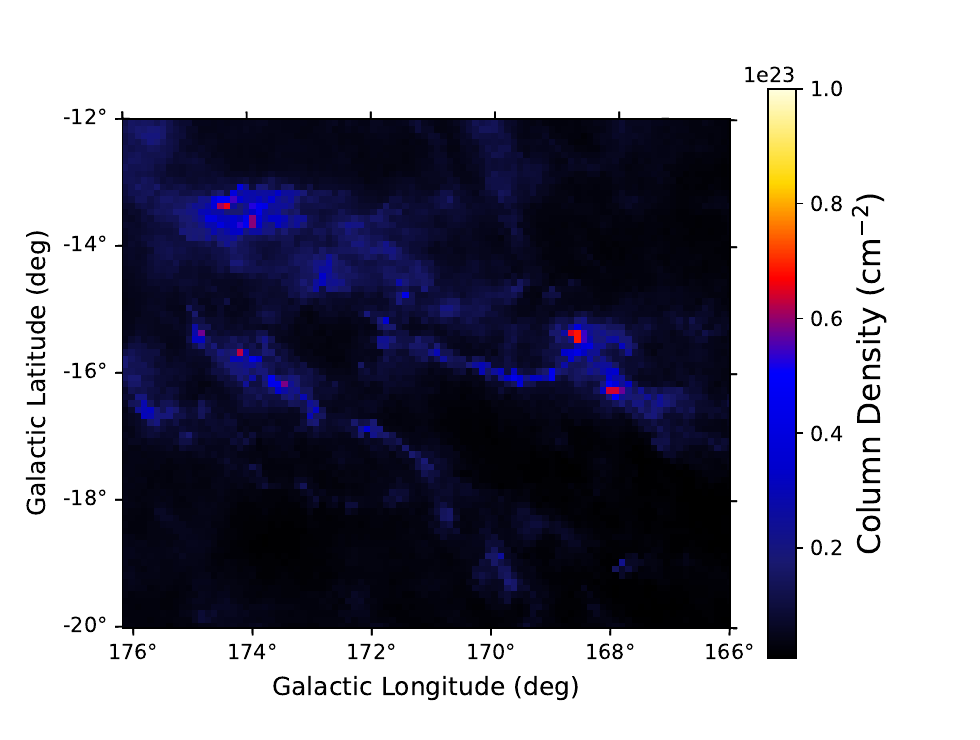}
        \caption{Taurus}
    \end{subfigure}

    \caption{The total column density (in units of cm$^{-2}$) of GMCs derived from the Plack dust opacity. }
    \label{fig:dustmap}
\end{figure*}
\clearpage
\begin{figure*}
    \centering

    \begin{subfigure}{0.35\linewidth}
        \includegraphics[width=\linewidth]{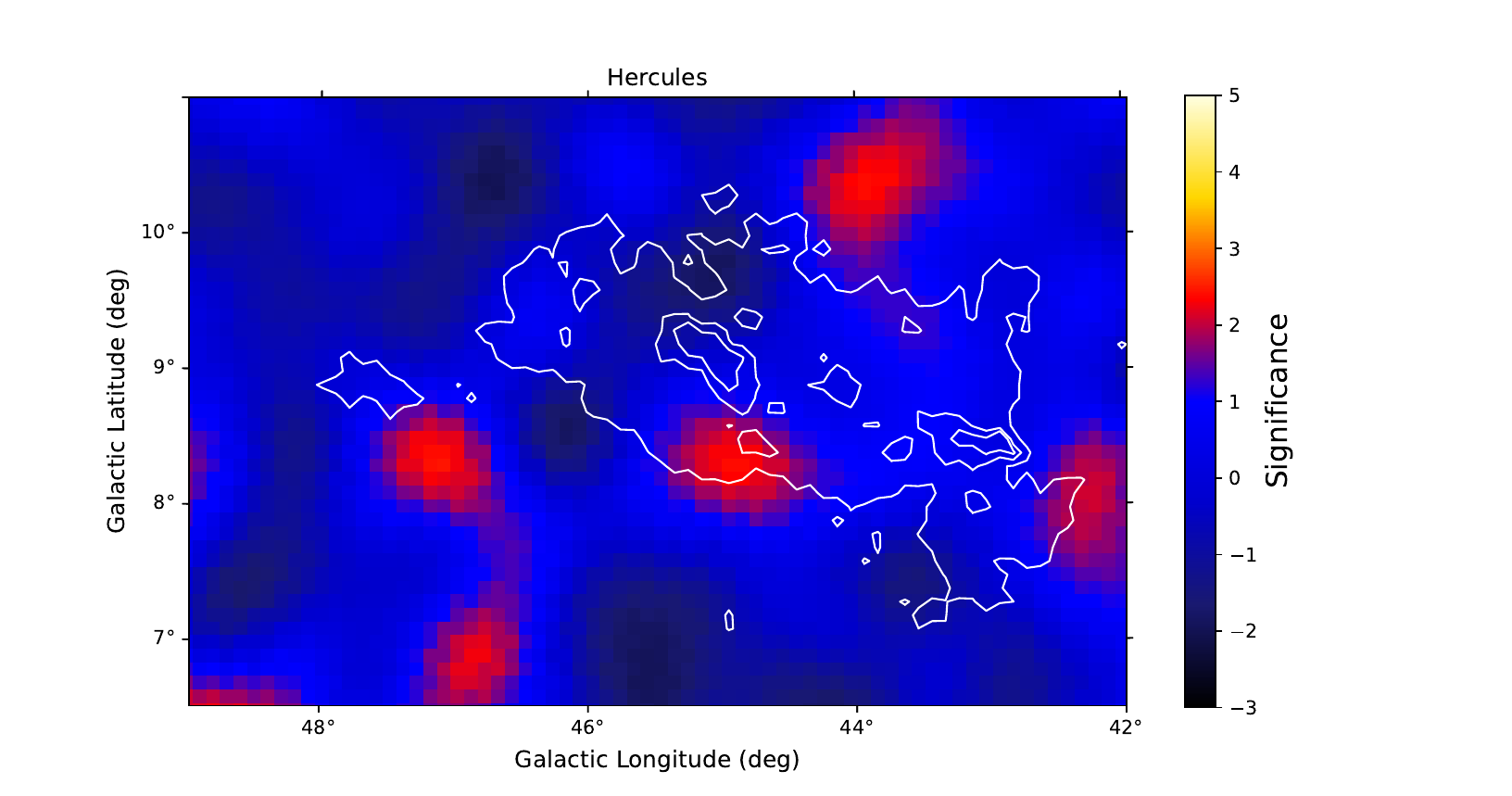}
        \caption{Hercules}
    \end{subfigure}
    \begin{subfigure}{0.35\linewidth}
        \includegraphics[width=\linewidth]{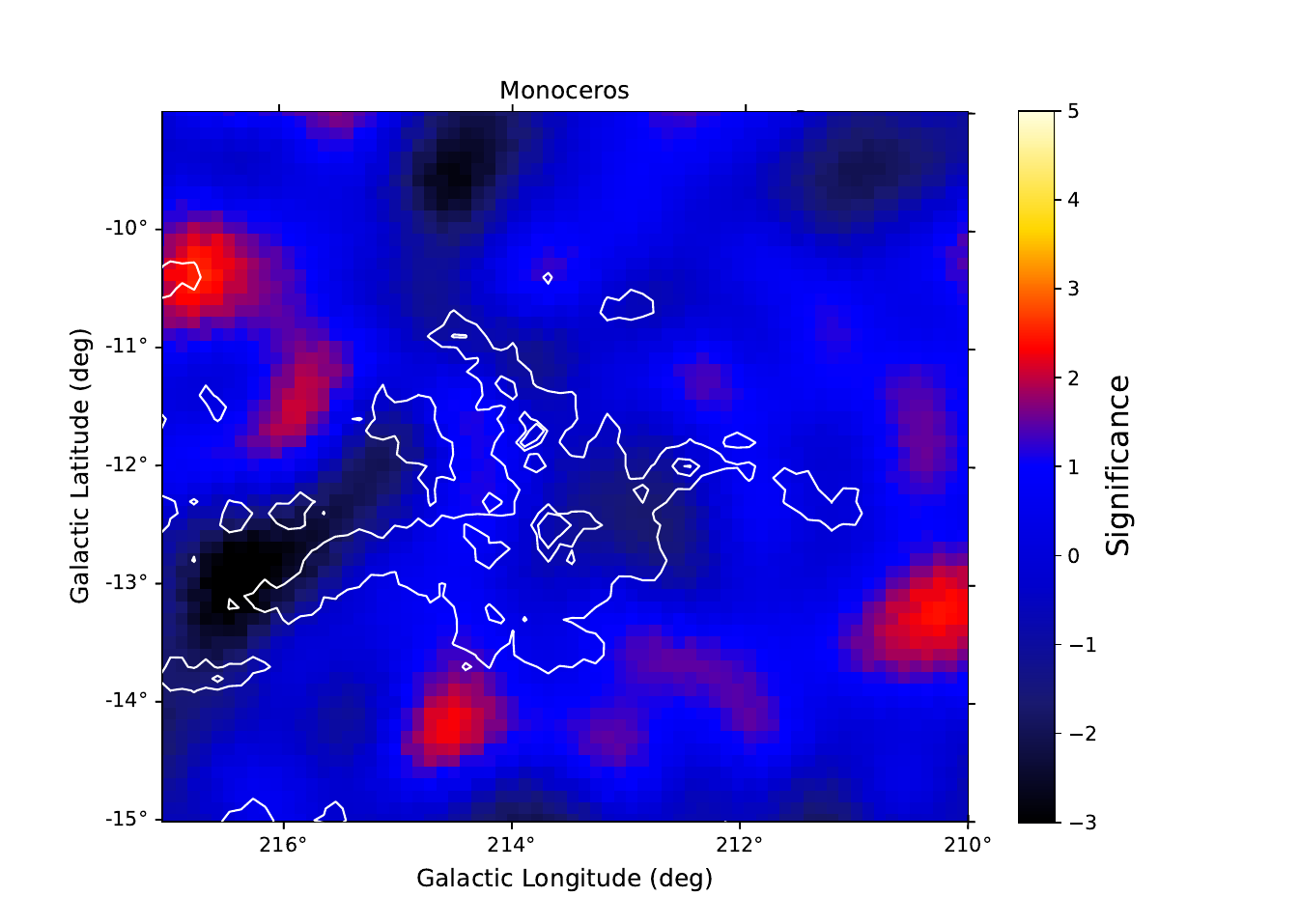}
        \caption{Monoceros}
    \end{subfigure}
    
    \begin{subfigure}{0.35\linewidth}
        \includegraphics[width=\linewidth]{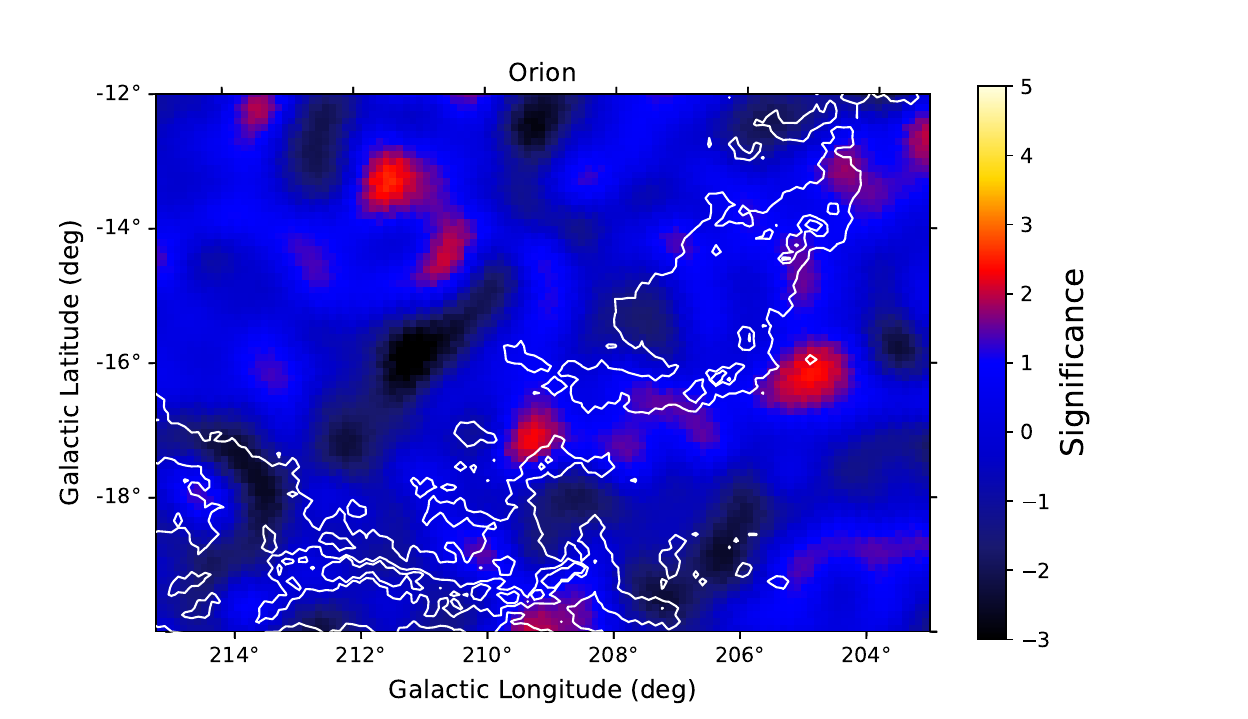}
        \caption{Orion}
    \end{subfigure}
    \begin{subfigure}{0.35\linewidth}
        \includegraphics[width=\linewidth]{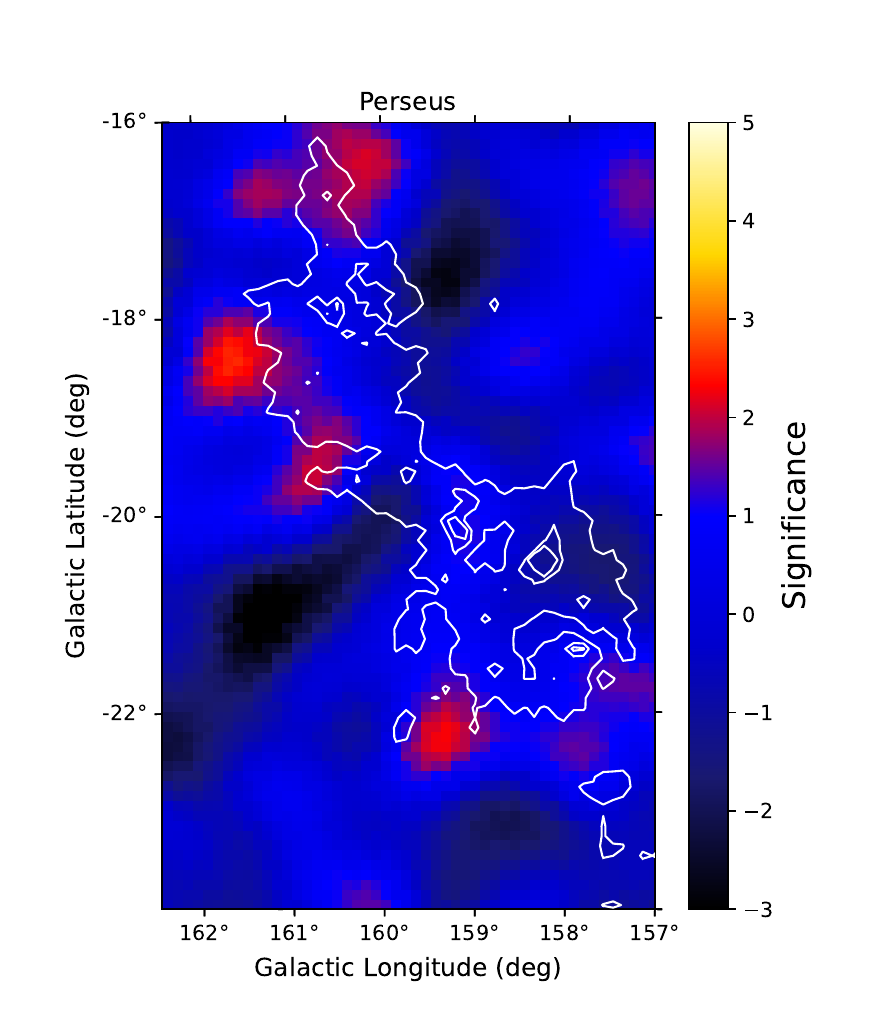}
        \caption{Perseus}
    \end{subfigure}
    
    \begin{subfigure}{0.35\linewidth}
        \includegraphics[width=\linewidth]{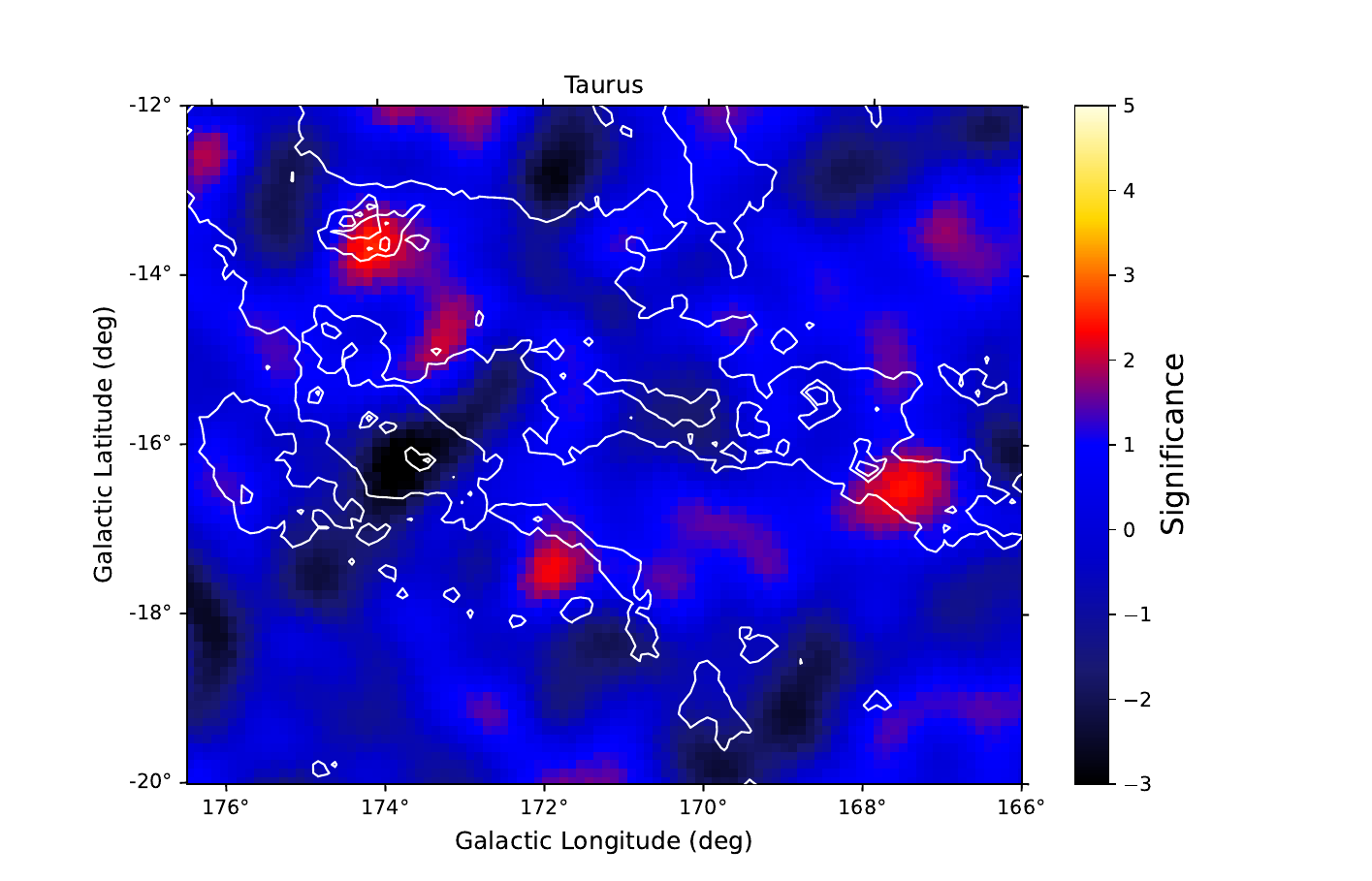}
        \caption{Taurus}
    \end{subfigure}

    \caption{The significance map of the GMC regions at energy above 25 TeV. The contours represent the regions in gas column density larger than 7 $\times$10$^{21}$ cm$^{-2}$ (for Hercules is 4 $\times$10$^{21}$ cm$^{-2})$.}
    \label{fig:km2a-skymap}
\end{figure*}

\clearpage
\begin{figure*}
    \centering
    \includegraphics[width=0.75\linewidth]{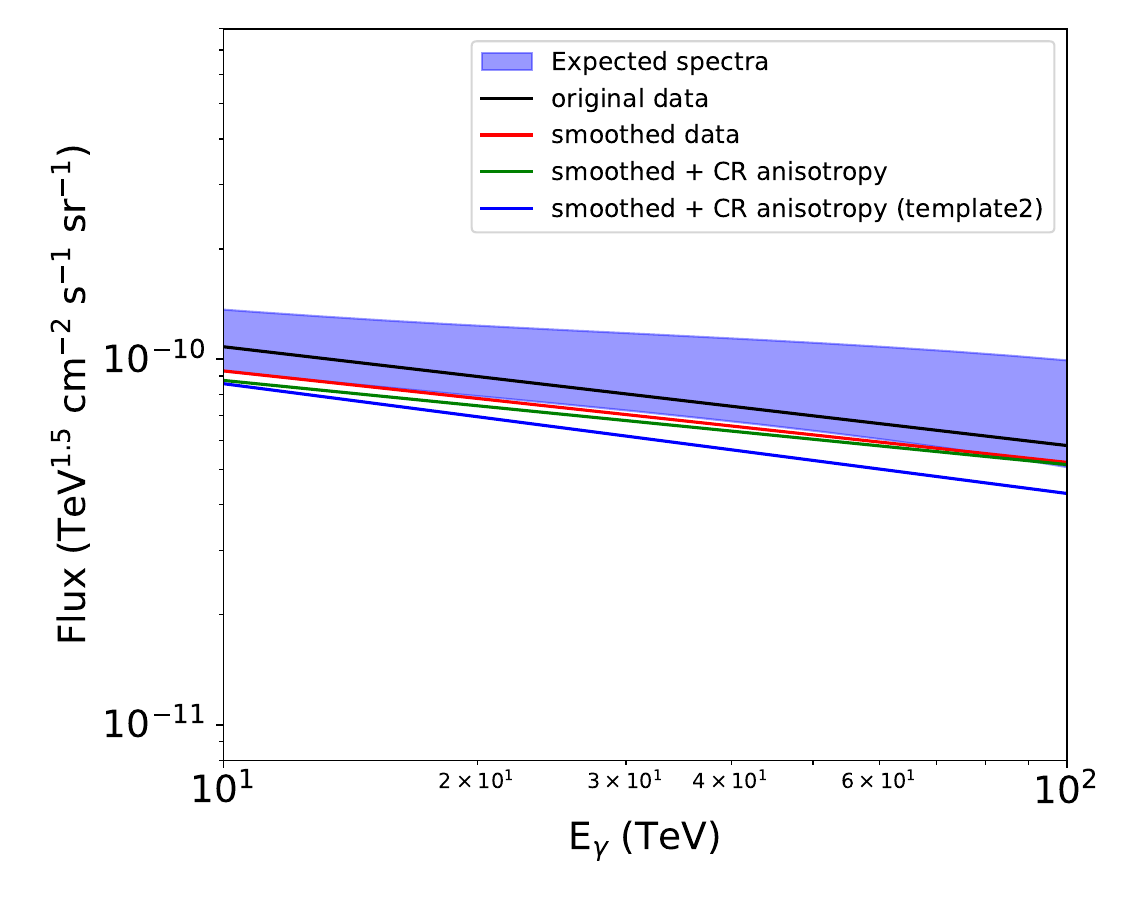} 
    \caption{The SED of the GMCs as measured by LHAASO-KM2A with different source templates and different datasets. }
    \label{fig:Differentdata}
\end{figure*}

\clearpage
\begin{figure*}
    \centering
    \includegraphics[width=0.45\linewidth]{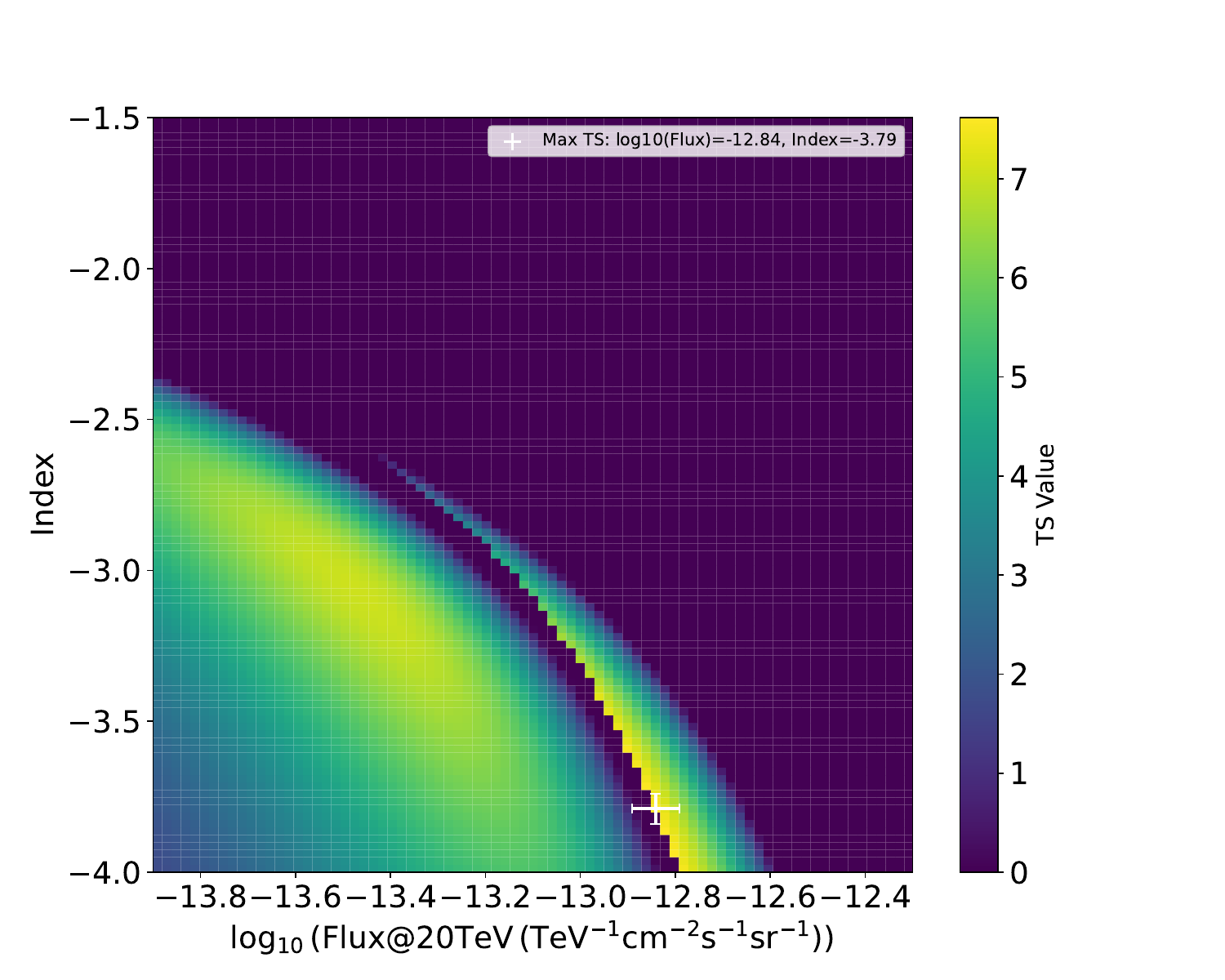}
    \includegraphics[width=0.45\linewidth]{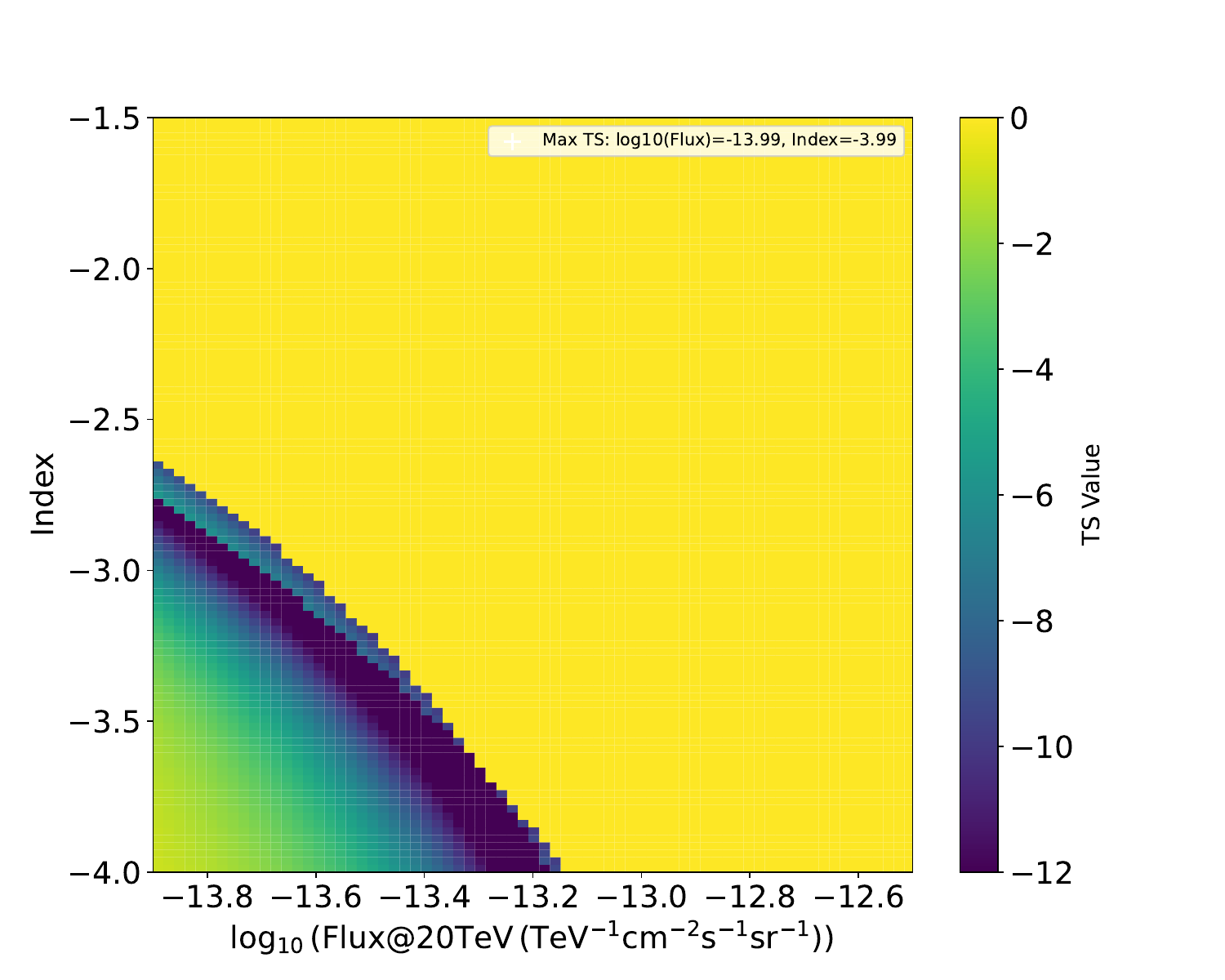}
    \caption{Left figure shows the TS profile for GMCs sample based on assumed the GMC follows a disk template using LHAASO-KM2A datasets. The right figure shows a stacked TS profile for randomly selected regions. The template selected was the same as used in this work. The max TS for both templates is less than 10, supporting the interpretation that the observed signal is correlated with the molecular cloud.}
    \label{fig:disk_template}
\end{figure*}

\begin{table}
    \centering
    \caption{ \bf The emission flux of the GMCs measured by WCDA and KM2A, together with 1$\sigma$ statistical. }
    \begin{tabular}{c c c}
     \hline
     \hline
     $<$E$>$    &  Flux &\\
     $<$TeV$>$ & (TeV cm$^{-2}$ s$^{-1}$ sr$^{-1}$)& \\
     \hline
     1.27 & (1.18 $\pm$ 0.50) $\times$ 10$^{-10}$& \\
     2.32 & (7.46$^{+4.59}_{-4.48}$) $\times$ 10$^{-11}$&\\
     3.90 & (6.59$^{+2.81}_{-2.79}$) $\times$ 10$^{-11}$&\\
     7.60 & (5.87$^{+2.14}_{-2.12}$) $\times$ 10$^{-11}$&\\
     12.59& 7.1 $\times$ 10$^{-11}$&Up limits\\
     19.95& 2.88 $\times$ 10$^{-11}$&Up limits\\
     31.62& (1.52$^{+0.75}_{-0.74}$) $\times$ 10$^{-11}$&\\
     50.12& (5.50$^{+4.16}_{-4.05}$) $\times$ 10$^{-12}$\\
     79.43& (6.11$^{+2.92}_{-2.59}$) $\times$ 10$^{-12}$&\\
     125.9& (4.28$^{+1.05}_{-0.88}$) $\times$ 10$^{-12}$&\\
     199.5& 4.57 $\times$ 10$^{-11}$&Up limits\\
     \hline
     \hline
    \end{tabular}
    
    \label{tab:sed_lhaaso}
\end{table}

\clearpage


\end{document}